\documentstyle[mncite,epsf]{mn}

\newcommand{\fig}{Figure}
\newcommand{\tab}{Table}
\newcommand{\sct}{Section}
\newcommand{\app}{Appendix}
\newcommand{\eqn}{Equation}

\newcommand{\heii}{\mbox{He\,\footnotesize II}}

\newcommand{\lsi}{$\mathrel{\hbox{\rlap{\hbox{\lower2pt\hbox{$\sim$}}}\raise2pt
\hbox{$<$}}}$}
\newcommand{\gsi}{$\mathrel{\hbox{\rlap{\hbox{\lower2pt\hbox{$\sim$}}}\raise2pt
\hbox{$>$}}}$}

\title[Modelling of the magnetic accretion flow in HU~Aquarii]
      {Modelling of the magnetic accretion flow in HU~Aquarii}

\author[Claus Heerlein, Keith Horne and Axel D. Schwope] 
       {Claus Heerlein$^{1,2,3}$, Keith Horne$^{1}$ and Axel D. Schwope$^{2}$ \\
       $^{1}$ University of St. Andrews, School of Physics and
              Astronomy, North Haugh, St. Andrews, Fife KY16 9SS, Scotland\\
       $^{2}$ Astrophysikalisches Institut Potsdam, An der Sternwarte
              16, D-14482 Potsdam, Germany\\ 
       $^{3}$ Institut f\"ur Theoretische Physik II,
              Universit\"at Erlangen-N\"urnberg, Staudstr. 7, D-91058 Erlangen, Germany\\
              \ \ e-mail: claus.heerlein@theorie2.physik.uni-erlangen.de  
}

\date{ 
submitted: 1998 May 5, re-submitted: 1998 September 18 
}
\pagerange{\pageref{firstpage}--\pageref{lastpage}}
\def\LaTeX{L\kern-.36em\raise.3ex\hbox{a}\kern-.15em
T\kern-.1667em\lower.7ex\hbox{E}\kern-.125emX}

\begin{document}
\maketitle
\label{firstpage}
\begin{abstract}
    We present a magnetic stripping model for AM~Her
    type objects. Our model is based on an equilibrium condition
    between ram pressure and magnetic pressure in a stiff dipolar
    magnetic field. We investigate the detailed geometry of the
    stripping process, most of which can be tackled
    analytically. By involving additional numerical calculations, the
    model allows the prediction of phase-resolved spectra and Doppler
    tomograms. The emission line features from the companion star, the
    horizontal stream and the accretion curtain are
    identified with the emission line components 
    found by Gaussian fitting to observational data of HU~Aqr in its
    high accretion state. Given the simplicity of the model,
    its agreement with the observation is remarkably good and enables 
    us to derive a number of physical quantities such as stellar masses 
    and radii, total mass accretion rate, bulk temperature, and coupling density.  
\end{abstract}


\section{Introduction}
    Magnetic cataclysmic variables (MCVs) are binary systems
    consisting of a late-type near main sequence secondary star and a
    magnetic white dwarf primary. Matter lost by Roche lobe overflow
    from the secondary star interacts with the strong magnetic field of the
    white dwarf before being finally accreted. These systems are ideal
    laboratories for studying the interaction of supersonic gas
    streams with strong magnetic fields.

    In an AM~Her object, or ``polar'', the magnetic field of the
    synchronously rotating white dwarf is sufficiently strong to
    prevent the formation of an accretion disk.  A relatively bright
    representative of this class is the eclipsing polar HU~Aqr with
    an orbital period of $P=0.\!^d08682$. It
    has been under intensive study (e.g. Glenn et al. 1994, Hakala 1995,
    Schwartz et al. 1995) since its identification in
    1992 (Hakala  et al. 1993, Schwope et al. 1993). The strength and 
    orientation of the primary's dipolar
    magnetic field has been determined with fair precision from
    cyclotron harmonics observed in low resolution spectra, from
    polarisation measurements, and from the X-ray light curve; it was 
    found for the magnetic field strength and
    the position of the accretion spot, respectively, $B_s \approx
    35\mbox{MG},\; \varphi_s \approx 40^o,\; \vartheta_s \approx
    26^o$. The mass ratio and inclination of the system remain
    somewhat more uncertain ($Q:=  M_{\mbox{\tiny wd}}/M_{\mbox{\tiny
    rd}}  \in [2.5,5.7]$, $i \in [81^o,90^o]$), but the
    eclipse phase width $\Delta\phi=0.0763$ however imposes a tight 
    relationship between $Q$ and $i$. Doppler maps generated from high
    resolution trailed spectra by Schwope et al. (1997a) clearly show that
    the illuminated face of the secondary star's Roche lobe and the
    accretion stream in the orbital plane are the dominant emission-line features of
    the high accretion state. The observed structure of the
    horizontal stream indicates that the stream initially 
    follows the ballistic trajectory and we are therefore dealing 
    with the situation described by Liebert \& Stockman (1985), in
    which thermal pressure dominates magnetic pressure at the inner 
    Lagrangian point $L_1$. Recent observations suggest that this  
    no longer holds for the reduced
    accretion state, when the density of the stream is significantly
    lower (Schwope et al. 1997b). Then, as proposed by Mukai (1988),
    the flow may couple directly to the magnetosphere at the
    $L_1$-point as assumed in early AM~Her models 
    (e.g.~Schneider \& Young (1980) for VV Pup).

\begin{figure}
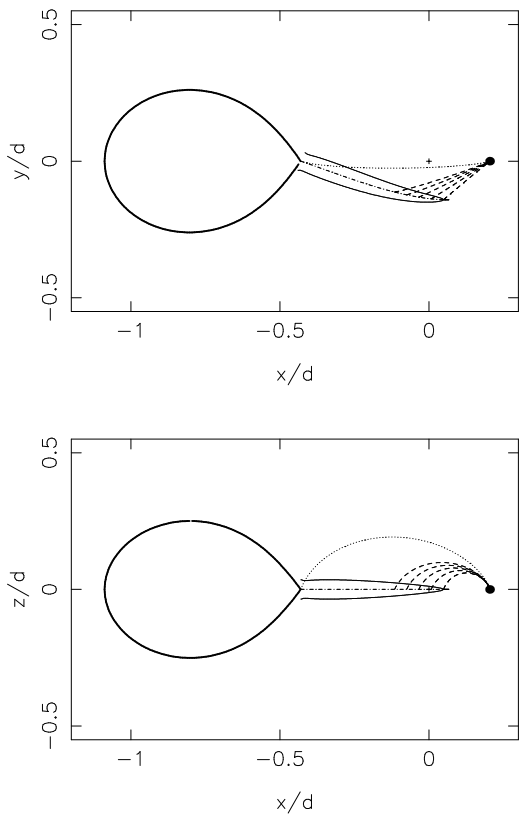
 
        \setlength{\unitlength}{1mm}
        \begin{picture}(70,120)
            \put(0,125){\includegraphics{images/fig1a.ps}}
            \put(0, 68){\includegraphics{images/fig1b.ps}} \noindent 
        \end{picture}
        \caption{Sketch of the HU~Aqr system projected on the
        x-y-plane (top) and x-z-plane (bottom); see text for details.}
        \label{geo_fig}  
    \end{figure}

    In this article we develop a simple but predictive model for phase-resolved
    emission line profiles and Doppler tomograms of MCV's. We test this model
    on observations of HU~Aqr in its high accretion state.  
    The accretion stream starts at the inner Lagrangian point $L_1$
    from the Roche lobe of the secondary star as shown in
    \fig~\ref{geo_fig}. The initial part of the stream,  not
    yet affected by the magnetic field, is confined to the orbital plane
    and is hence called the horizontal stream. In our calculations it 
    is assumed to have
    a Gaussian density profile that is however truncated when the 
    density drops below a critical value below which the 
    matter couples to the
    magnetosphere. This stripping threshold, displayed by the solid
    lines around the stream centre in \fig~\ref{geo_fig}, is defined
    by the ram pressure in the stream dropping below the magnetic
    pressure of a stiff dipole field. The stripped material, as it 
    leaves the orbital plane, forms a two-dimensional accretion curtain
    and glides along the field lines until it is finally accreted on
    the white dwarf's surface. Even though magnetic stripping occurs
    virtually over the whole length of the horizontal stream, most of the
    matter is stripped near the end of the horizontal stream,
    as indicated by the  dashed lines in \fig~\ref{geo_fig}. 
    The curtain trajectory passing through the end point of the 
    horizontal stream has been named vertical stream in the literature.
    The dotted line marks the trajectory of matter
    that is stripped  right at the $L_1$-point (see below).

    X-ray radiation from the accretion region excites line emission
    from the companion star surface and the accretion stream. The line
    emission from the secondary star was found to be optically thick
    in our initial study (Schwope et al. 1997a), whereas for the
    emission from the horizontal stream and the curtain the situation
    is less obvious and we are here comparing the limiting cases of
    optically thin and thick radiation reprocessing. 

    We first focus in \sct~\ref{magn_str} on the horizontal stream and
    derive the stripping rate with which matter is removed from it by
    the magnetic field. In \sct~\ref{accr_curt} we investigate the
    geometry of the accretion curtain in which the matter propagates
    through the magnetosphere. Our model provides the velocity vector and 
    densities at each point of the flow, enabling us to model phase-resolved 
    spectra and Doppler maps in  \sct~\ref{spect}, which are fitted to
    observations of HU~Aqr in \sct~\ref{fit_sct}. This allows us to derive
    the physical quantities of the system such as stellar masses and radii, total
    mass accretion rate, bulk temperature, and coupling density. 
    Finally in \sct~\ref{discus} we discuss the discrepancies of model and observation 
    and propose further improvements.

\section{Magnetic stripping}
    \label{magn_str}
\subsection{The horizontal stream}
    As suggested by the high-state observations, the stream is not
    strongly affected by the magnetic field as it passes the inner
    Lagrangian point $L_1$. This situation invites use of the
    semi-analytical solution for an inviscid isothermal flow in a
    Roche potential given by Lubow \& Shu (1975, 1976) with additional
    corrections for the magnetic influence. First the applicability of
    the Lubow-Shu-model on AM~Her stars deserves some discussion:
    (a) The stream and secondary star are heated by ionising radiation from the accretion
    region. The diminished line emission of the secondary's leading side implies
    considerable shielding by the accretion curtain (Schwope 1997a),
    which must therefore have a high optical depth for the ionising
    radiation. The column density there is even lower than in the
    horizontal stream, where we consequently also expect a high
    optical depth. This confines the irradiative heating to the skin
    layers and causes a highly inhomogeneous temperature distribution
    over the stream. We assume that the bulk of matter in the stream
    centre remains at a constant temperature and the overall behaviour
    of the stream is not much affected by the hot surface layer. We
    also neglect the impact of radiation pressure on the stream.
    (b) In the region where the magnetic pressure already overcomes
    the thermal pressure but not yet the ram pressure, which is the
    case for almost the whole stream apart from a very thin shell
    around the $L_1$-point, instabilities shatter the stream to small
    blobs (Liebert \& Stockman 1985). We assume that the mean density
    on a scale comparable to the stream width is
    still described by a simple Gaussian profile as originally
    proposed by Lubow \& Shu. In the Reynolds averaged sense we treat
    the ensemble of gas blobs as a quasi-steady flow and assume a
    vanishing eddy-viscosity therein. 
    (c) Further downstream magnetic stripping erodes the stream
    from the outside. This  violates the assumption of a Gaussian
    density distribution and might also affect the temperature
    structure. Even though the stream motion is highly supersonic the
    internal gas pressure is important in the transverse directions of
    the flow. Our model however assumes for simplicity that the
    magnetic pressure exerted on the surface of the stream is equal 
    to the pressure that the stripped layers would have exerted, so that the
    kinematic behaviour of the core is not affected by the missing envelope.

    Following the suggestion of Lubow \& Shu (1975) we formulate the
    problem in dimensionless units of length, time and mass:

    \begin{equation}
        1[L] = d \; , \quad
        1[T] = \Omega^{-1} \; , \quad   
        1[M] = \dot{M}\Omega^{-1} \; ,
        \label{units}
    \end{equation}
    where $d$ is the orbital separation of the system, $P=2\pi/\Omega$
    its rotation period, and $\dot{M}$ the total mass accretion rate.

    Also from Lubow \& Shu (1975) we adopt the treatment as a
    perturbation analysis in a small parameter $\epsilon$, which is
    the isothermal sound speed $\sqrt{kT/m}$ in the stream for a
    temperature $T$ and a mean molecular weight $m$ in above units
    given by \eqn~(\ref{units}). Under the assumption of an isothermal flow the
    perturbation parameter $\epsilon$ is constant over the whole
    accretion stream. For the orbital period of the HU~Aqr system we
    have in case of a pure hydrogen gas $\epsilon = 0.011 (T/10^4K)^{1/2}
    (d/10^9m)^{-1}$.

    Initially we consider a stream without a magnetic field. We use
    the natural coordinate system of the horizontal stream where $s$
    measures the distance from the $L_1$-point along the stream centre
    and the pair $(n, z)$ represents the distances from the stream
    centre parallel and perpendicular to the orbital plane. Lubow \&
    Shu (1975, 1976) propose that the density in the horizontal stream
    has a Gaussian distribution around the ballistic trajectory, i.e.
   \begin{equation}
        \rho(s,n,z) = \rho_{0}(s) \; \exp(-\frac{1}{2\epsilon^2}(\gamma
           n^2 + \chi z^2)) \; ,
      \label{rho_exp}
    \end{equation}
    where the characteristic horizontal and vertical widths
    $1/\sqrt{\gamma}$ and $1/\sqrt{\chi}$, respectively, are functions
    of $s$ only and can be found from an initial value problem as 
    the solution of a closed set of non-linear singular differential
    equations, in which the mass fraction $\mu$ appears as a
    parameter; the mass fraction is defined as the white dwarf mass
    over the total mass: $\mu := M_{\mbox{\tiny wd}}/(M_{\mbox{\tiny
    wd}}+M_{\mbox{\tiny rd}})$. A proper numerical integration of
    these equations requires asymptotic analysis near the singular
    point $L_1$ (see \app~\ref{series_app}).
    With $u_{s0}$ as the s-component of the stream velocity in the
    centre we have in lowest order of $\epsilon$
    \begin{equation}
        \rho_{0}(s) = \frac{\sqrt{\gamma\chi}}{2\pi \epsilon^2 u_{s0}} \; ,
        \label{normalize}
    \end{equation}
    since we have normalised our equations to the total mass flow, i.e. 
    $  \int\rho \, u_{s0} \;dn\,dz = 1 $
    and the local stream velocity $u_s$ can be approximated by the 
    central velocity $u_{s0}$ in lowest order of $\epsilon$. 

\subsection{The stripping process}\label{subsec-stripping}
    The ram pressure of the stream orthogonal to the magnetic field 
    can be written as 
    \begin{eqnarray}
        p_{\mbox{\tiny ram}} &=& \frac{1}{2} \,\rho u_{s0}^2
            \sin^2\varpi = \nonumber \\ &=& \frac{\sqrt{\gamma \chi} \;
            u_{s0}\sin^2\varpi}{4 \pi \epsilon^2 } \; 
            \exp(-\frac{1}{2\epsilon^2 }(\gamma n^2 + \chi z^2)) \; ,
        \label{p_ram}
    \end{eqnarray}
    where $\varpi$ is the stripping angle between stream motion and
    the magnetic field. Further the thermal pressure of the gas is given by 
    \begin{equation}
        p_{\mbox{\tiny th}} =  \epsilon^2 \, \rho =
        \frac{\sqrt{\gamma\chi}}{2\pi u_{s0}} \;
        \exp(-\frac{1}{2\epsilon^2 }(\gamma n^2 + \chi z^2))\; . 
        \label{p_therm}
    \end{equation}
    We now compare the total gas pressure $p_{\mbox{\tiny th}}+
    p_{\mbox{\tiny ram}}$ in the stream with the local magnetic
    pressure $ p_{\mbox{\tiny mag}} = B^2/(8 \pi) $. We assume then
    that the stripping threshold is given by  
    $p_{\mbox{\tiny th}}+p_{\mbox{\tiny ram}}=p_{\mbox{\tiny mag}}$.
    Since we are considering the stripping process at large distances
    from the dipole source we approximate the local $B$ by the magnetic 
    field $B_0$ at the stream centre with an error of order $\epsilon$.
    Using \eqn~(\ref{p_ram}) \&
    (\ref{p_therm}) we find that the critical surface is a
    centred ellipse with the principal axes $a_n$ and $a_z$ aligned to
    the coordinate axes perpendicular to the direction of propagation:
    \begin{equation}
        a_{n} =\epsilon \, \sqrt{\frac{2 \sigma}{\gamma}}\;, \quad
        a_{z} =\epsilon \, \sqrt{\frac{2 \sigma}{\chi}}  
        \label{thres}
    \end{equation}
    with
    \begin{equation}
        \sigma := \ln \Big ( {\frac{2  \sqrt{\gamma
        \chi}\,(u_{s0}^2\sin^2\varpi+2\epsilon^2)}
        {\epsilon^2 B_{0}^2 \, u_{s0}}} \Big ) \; .
        \label{sigma}
    \end{equation}

    \begin{figure}
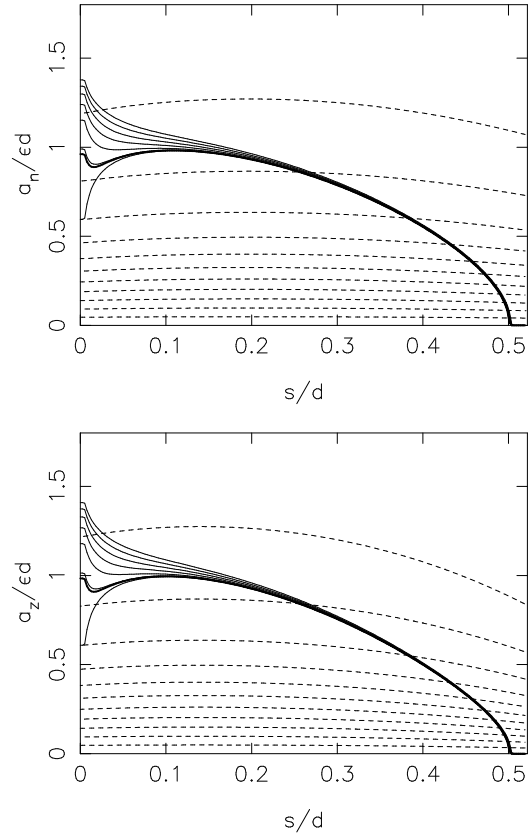

        \setlength{\unitlength}{1mm}
        \begin{picture}(75,120)
            \put(0,125){\includegraphics{images/fig2a.ps}}
            \put(0, 68){\includegraphics{images/fig2b.ps}}
        \end{picture}
        \caption{Stripping threshold (solid lines) in n and z
        direction, respectively for $\epsilon = 0$ to $0.24$ in steps of $0.04$; 
        the remaining parameters are as in \tab~\ref{param_tab}.
        Superimposed (dashed lines) are  stream lines
        for the 10\% to 90\% (in steps of 10\%), 97.5\% and 99.9\%
        percentiles.}
        \label{thres_fig}
    \end{figure}

\begin{figure}
    \setlength{\unitlength}{1mm}
        \begin{picture}(70,60)
            \put(0,66){\includegraphics{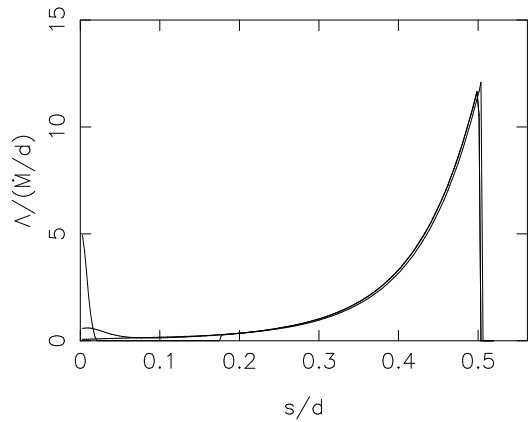}}
        \end{picture}
        \caption{ The stripping rate $\Lambda$ according to
                \eqn~(\ref{lambda}) for the HU~Aqr geometry
                (see \tab~\ref{param_tab}); $\epsilon$ takes the values
                0.04, 0.12, and 0.36.
                }
        \label{sr_fig}
    \end{figure}

    For numerical calculations we assume a dipole field
    \begin{equation}
        {\bf B}=\frac{3\, {\bf r}_{\mbox{\tiny wd}}
             ({\bf r}_{\mbox{\tiny wd}}\cdot{\boldmath \cal M}) 
             - r_{\mbox{\tiny wd}}^2{\boldmath  \cal M}}
             {r_{\mbox{\tiny wd}}^5}\;,
    \end{equation}
    with $r_{\mbox{\tiny wd}}$ the distance from the white dwarf
    centre and ${\bf \cal M}$ the magnetic
    moment. \fig~\ref{thres_fig} shows numerically calculated
    streamlines and stripping surfaces for the geometry of
    HU~Aqr. Minima in  $a_n$ and
    $a_z$ occur immediately after the $L_1$ point for 
    $\epsilon \, {\stackrel{\displaystyle\displaystyle<}{_{_{\displaystyle \sim}}}} 0.08$, i.e. for a cold stream. They arise because the thermal
    pressure decreases due to the dilution caused by the acceleration
    of the stream before the ram pressure increases enough to take over the
    pressure balance with $p_{\mbox{\tiny mag}}$ that provides our
    stripping condition. The locations of the minima are close to the
    equilibrium point between thermal pressure and magnetic pressure,
    downstream from which Liebert \& Stockman (1985) postulate
    shattering of the stream into fine blobs. The presence of minima
    in the stripping thresholds suggests a `magnetic nozzle' at the
    beginning of the horizontal stream, however one should be cautious
    with that interpretation since the diamagnetism of the secondary
    star, and possibly its intrinsic magnetic field, induce in the
    vicinity of $L_1$ a large distortion of the ideal dipole field
    assumed in our model. The observed trailed spectra of HU~Aqr show
    no evidence for a manifest stripping directly at the $L_1$-point. 
    Cropper (1986) however concludes from polarisation measurements in 
    ST~LMi that a two-spot accretion is the best fit to his data, so
    stripping directly at the $L_1$-point may occur in other AM~Her 
    objects. Since we can fit the
    parameter $\epsilon$ only very indirectly
    (c.f.~\sct~\ref{fit_sct}) it is very reassuring that towards the
    end of the horizontal stream, where we expect the largest
    proportion of the magnetic stripping to occur (c.f.~\fig~\ref{sr_fig}), 
    the solution becomes virtually independent of $\epsilon$.

    Neglecting diffusion at the
    edges of the stream, the shape of a
    truncated Gaussian density profile is preserved as
    the stream propagates. The maximum mass flow that
    can survive magnetic stripping and proceed beyond 
    position $s$ on the horizontal stream is given by
    $   \int_{(\frac{n}{a_{n}})^2+(\frac{z}{a_{z}})^2\;\le\;1} \; 
        \rho \, u_{s0} \; dn\,dz \; .
    $
    Consistent with neglecting diffusion, we consider only the mass 
    flow $\dot{m}$ on those stream lines that have not yet coupled to the
    magnetosphere. This introduces a minimum over all previous stream
    coordinates $s$. This has a significant effect only for  
    $\epsilon \, {\stackrel{\displaystyle<}{_{_{\displaystyle \sim}}}} 0.08$,
    for which stripping near $L_1$ occurs. 
    Evaluation of the integral in elliptical coordinates
    and substitution of \eqn~(\ref{rho_exp}),  (\ref{normalize}) and
    (\ref{thres})   results finally in 
    \begin{eqnarray}
        \dot{m}(s) &=& \min_{s'\le s} \Big(1-e^{-\sigma(s')} \Big) 
             = \nonumber \\ &=&  1\,-\, 
             \max_{s'\le s} \Big( \frac{\epsilon^2 B_{0}^2 \, u_{s0}}{2
             \,\sqrt{\gamma \chi}\,(u_{s0}^2\sin^2\varpi+2\epsilon^2)\,
             } \Big) \; .
        \label{mdot}
    \end{eqnarray}
    The symmetry of the integration region causes the first order error terms 
    to cancel. The resulting stripping rate $\Lambda = -d\dot{m}/ds$ is then 
    given by
    \begin{equation}
        \Lambda = \frac{d}{ds} \;\; \max_{s'\le s} \Big( \frac{\epsilon^2 B_{0}^2
        \, u_{s0}}{2 \,\sqrt{\gamma
        \chi}\,(u_{s0}^2\sin^2\varpi+2\epsilon^2)} \Big) \; ,
    \label{lambda}
    \end{equation}
    wherein $B_{0}$, $u_{s0}$, $\gamma$ and $\chi$ are all regarded
    as functions of~$s$. 

    At the $L_1$-point we have a mass flow $\dot{m}$ slightly below 1
    since the magnetic pressure is very small in comparison with the
    gas pressure there but it is not vanishing totally. This is not a
    major inconsistency in our model since, as discussed  above, our
    approximation of a dipole field is already very inaccurate in this
    region. However for sensible values of $\epsilon$ the stripping
    occurs mainly at large distances from the $L_1$-point, where our
    approximation holds well (see \fig~\ref{sr_fig}). The allowed mass flow
    $\dot{m}$ is monotonically decreasing by construction of
    \eqn~(\ref{mdot}), so un-physical negative stripping rates are
    ruled out. The above calculation did not account for blocking of
    the stripped matter by the dense core of the stream as proposed by
    Mukai (1988). Surface currents induced in the stream lead to a
    large local distortion of the threading field  which then
    voids the stream; hence  our model assumes that matter stripped 
    of the back side is led around the stream by the distorted field lines.

\section{Geometry of the accretion curtain}
    \label{accr_curt}
\subsection{The threading process}
    Liebert \& Stockman (1985) point out that the threading of matter
    onto the magnetic field involves a variety of physical processes,
    which are not yet fully understood (c.f. Cropper 1990, Warner 1995). 
    However it seems clear that in accretion onto compact objects magnetic
    instabilities play the key role. Burnard et al. (1983) have shown
    that Kelvin-Helmholtz instabilities caused by a shear flow with
    inhomogeneous density distribution generate small drops which are
    easily captured by the magnetic field.  It is also suggested that,
    even though the gas stream is highly supersonic, the negative density 
    gradients additionally produce Rayleigh-Taylor instabilities. By these
    instabilities the stream is broken up into large diamagnetic
    blobs that, in contrast, are only slowly penetrated by the
    magnetic field.  We assume here that the stripping is primarily
    driven by Kelvin-Helmholtz instabilities. In our model the droplets stripped
    from the surface follow abruptly the magnetic field lines and
    their kinetic energy from the motion perpendicular to the field is
    instantaneously thermalised. It turns out that the resulting rise
    in temperature allows the flow to become subsonic again.

\subsection{The magnetic curtain}
    In the previous section we have modelled the rate at which material 
    from the horizontal stream is captured by the magnetic field. The
    geometry of the 
    threading region depends on the 
    mass ratio     $Q$ and on the
    orientation of the magnetic dipole $(\vartheta_d, \varphi_d)$
    relative to the canonical Cartesian system of the co-rotating
    frame. In polar coordinates $(r,\theta,\psi)$ aligned with the
    dipole we use the usual parameterisation of the field lines in
    azimuth $\psi$ and equatorial radius $r_m$; a dipole field line is
    then described as a function of the co-latitude $\theta$ by
    \begin{equation}
        \psi=\mbox{const}\;,\quad r=r_m \sin^2\theta \; .
        \label{Bpara}
    \end{equation}   

    For an arbitrary threading point $(x,y,0)$ on the horizontal stream
    we now consider the spherical triangle with the dipole axis, its
    projection onto the $x$-$y$ plane, and the threading point as
    corners. The angle $\varphi_c$ from the projected dipole axis to
    the threading point is given implicitly by
    \begin{equation}
        \tan(\varphi_d+\varphi_c) = \frac{-y}{1-\mu-x} \; ,
        \label{phi_c}
    \end{equation}
    wherein the quadrant is provided by the standard sign
    con\-ven\-tion for the denominator. Spherical
    tri\-go\-no\-me\-try provides the coupling co-latitude $\theta_{c}$
    and azi\-muth $\varphi_c$ in dipole coordinates
    \begin{equation}
        \cos \theta_{c} = 
             \cos \varphi_c \sin\theta_d\;,\quad \sin \psi = 
             \frac{\sin \varphi_c}{\sin\theta_c} \; ,
    \end{equation}   
    which is unique since in our case always $\theta_c \in [0,\pi[$
    and $\psi \in [-\frac{\pi}{2},\frac{\pi}{2}[$, it is also clear that always
    $|\sin \varphi_c| \le |\sin\theta_c|$.  
    The above set of
    equations seems to us slightly easier than the one given by
    Ferrario (1989), but nevertheless describes the same
    geometry. \eqn~(\ref{Bpara})  finally yields the equatorial radius
    of the threading field line
    \begin{equation}
        r_m = \frac{r_c}{\sin^2\theta_c} \; .
        \label{r_m}
    \end{equation}   

    We obtain the initial velocity with which the stripped material
    moves on the curtain from the purely kinematic argument that the
    component parallel to the field lines is conserved. For the dipole
    orientation in HU~Aqr (see \tab~\ref{param_tab}) it
    turns out that the material turns a rather sharp corner of $\varpi
    \approx 75^o$. If we use a mass specific Roche potential
    normalised with an additive constant such that $\Phi(L_1)=0$ the dissipated energy per unity
    mass is given as $ -\Phi_c\sin^2 \varpi$, where $\Phi_c$ is the
    potential at the coupling point. Towards the end of the horizontal
    stream the energy loss for an H-atom is nearly 1~keV.  
    We assume for our model that the resulting rise in temperature is relaxed
    instantaneously to the isothermal equilibrium temperature characterised
    by the isothermal sound speed $\epsilon$ - in practice a different value 
    of $\epsilon$ may apply to each field line in the curtain.

    Continuity and momentum conservation for an isothermal flow inside
    a flux tube provide a differential equation for the velocity $u$ along
    the field lines, given the magnetic field $B$ and the Roche potential $\Phi$
    \begin{equation}
        \frac{du}{u} \;\Big(\frac{\epsilon^2}{u^2}-1\Big) = 
            \frac{\epsilon^2}{u^2} \;\frac{dB}{B} + \frac{1}{u^2} \;
            d\Phi \; ; 
        \label{upde}
    \end{equation}   
    Shortly after coupling the stream becomes highly supersonic
    again, hence pressure effects are negligible everywhere and we get
    by integration
    \begin{equation}
        u \approx \sqrt{2(\Phi_c \sin^2\varpi - \Phi)}\; ,
    \end{equation}
    which is simply energy conservation for a free falling
    particle. In principle the field lines could lead so far uphill in
    the Roche potential that \eqn~(\ref{upde}) predicts the material to
    stagnate somewhere with zero velocity. For the HU~Aqr geometry
    this case never occurs in our simulation.

   \begin{figure}
        \setlength{\unitlength}{1mm}
        \begin{picture}(70,50)
            \put(8,0){\includegraphics{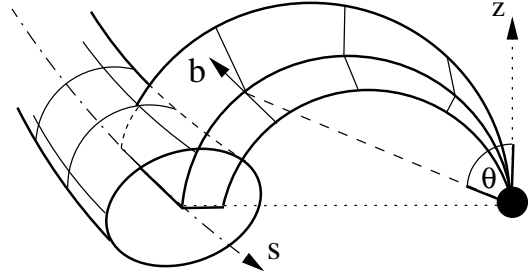}}
        \end{picture}
        \caption{Sketch of the horizontal stream with a part of the accretion curtain, 
                 which illustrates the coordinates introduced in the text.}
        \label{coords_fig}
    \end{figure}

    Gliding down the field lines the material from the stream 
    forms an essentially two-dimensional curtain.  We now introduce a curtain
    coordinate system (c.f.~\fig~\ref{coords_fig}) which consists of the distance $s$ of the
    coupling point from $L_1$, the magnetic co-latitude $\theta$, and
    the distance $b$ perpendicular to the curtain. The coordinates 
    are orthogonal at the coupling point and the
    curves of constant $b$ are field lines. Since the curtain flow 
    follows field lines, we can represent the density as
    \begin{equation}
        \rho = \sigma_c(s,\theta) \, f(\Delta^{-1}b) \; ,
    \end{equation}
    where the typical width $\Delta$ of the curtain depends on $s$ and
    $\theta$.
    The convergence of  field lines along the curtain flow 
    can be expressed by
    the variation of the thickness of a flux tube perpendicular to the
    direction of motion. For the flux tube starting at  stream
    coordinate $s$ the distance $dl$ to the adjacent field line starting at 
    $s+ds$ is given by 

    \begin{equation}
        \Big(\frac{\partial l}{\partial s}\Big)^2 
             =  h^2_{r_m} \Big(\frac{\partial r_m}{\partial s}\Big)^2 
               + h^2_{\psi}\Big(\frac{\partial \psi}{\partial s}\Big)^2  ,
    \end{equation}   
    where the derivatives $({\partial r_m}/{\partial s})$ and
    $({\partial \psi}/{\partial s})$ are a direct consequence of
    \eqn~(\ref{phi_c})-(\ref{r_m}) with $x=x(s)$ and $y=y(s)$. The metric coefficients
    $h^2_{r_m}$ and $h^2_{\psi}$ are given by \eqn~(\ref{h_r_m}) and
    (\ref{h_psi}) as functions of $\theta$ and $r_m$.

    The mass flow transported on the flux tube defined by
    $s$ and $s+d s$ is given by $\Lambda \, ds$ 
    (c.f.  \eqn~(\ref{lambda})). As a consequence of continuity we have
    then
    \begin{equation}
      \sigma_c = \frac{\Lambda}{u}\Big(\frac{\partial s}{\partial
             l}\Big)  = \frac{\Lambda}{u} \Big[h^2_{r_m}
             (\frac{\partial r_m}{\partial s})^2 +  h^2_{\psi}
             (\frac{\partial \psi}{\partial s})^2 \Big]^{-\frac{1}{2}};
    \end{equation}
    at the coupling point we know from simple geometry that
    $({\partial l}/{\partial s})_0 = \sec \varpi$ holds.
    Since the flow is bound to a flux tube, we can write the 
    typical width of the curtain as 
    $\Delta^{-1} \propto B ({\partial l}/{\partial s})$.
    A quite suitable choice for $f$ could be a Gaussian profile with
    a dispersion $\Delta$, for which at the coupling point 
    $\Delta^2=a_n a_z$ holds.

\section{Simulation of Doppler maps and trailed spectra}
    \label{spect}
    \begin{figure*}
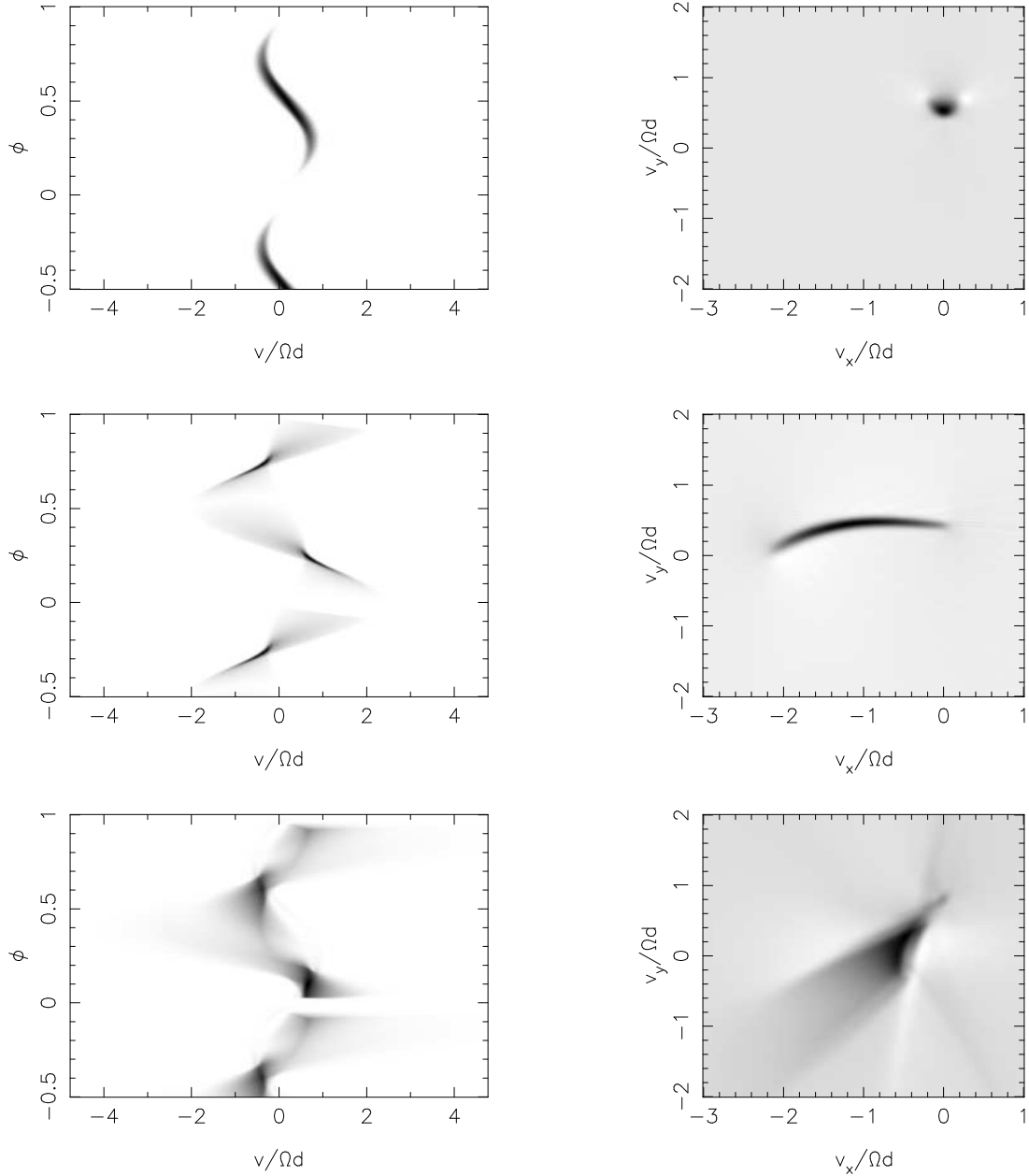

        \setlength{\unitlength}{1mm}
        \begin{picture}(170,170)
            \put(0,180){\includegraphics{images/fig4a.ps}}
            \put(90,180){\includegraphics{images/fig4b.ps}}
            \put(0,122){\includegraphics{images/fig4c.ps}}
            \put(90,122){\includegraphics{images/fig4d.ps}}
            \put(0,65){\includegraphics{images/fig4e.ps}}
            \put(90,65){\includegraphics{images/fig4f.ps}}
        \end{picture}
        \caption{Model line emission from various parts of the system
        displayed as trailed spectra (first column) and in the Doppler
        tomograms (second column) calculated by a filtered back-projection.  
        Rows: (a) companion Roche lobe, (b)
        horizontal steam, (c) accretion curtain. The gray-scale of each row
        is chosen separately. 
        Model parameters as given in  \tab~\ref{param_tab}.}
        \label{parts_fig}
    \end{figure*}

\subsection{Companion star line emission}

    Our model assumes that the companion star exactly fills its Roche
    lobe, that the reprocessed radiation is proportional to the
    absorbed X-rays, and that the emission is diffuse according to the
    foreshortening given by Lambert's law, i.e. with a  cosine
    dependency of the viewing angle. Isotropic re-emission has shown
    to provide a less accurate line profile. Thermal broadening at the
    bulk temperature causes a velocity  dispersion 
    $\epsilon\Omega d/c$, which is just above the numerical resolution
    of our spectral grid.  The X-rays are assumed to originate from a
    point source at the white dwarf's centre. The involved error in
    the location of the X-ray source is of the order of the white
    dwarf radius $R_{\mbox{\tiny wd}}$, i.e. typically 1\% of the
    orbital separation $d$.

    The upper panel of \fig~\ref{parts_fig} shows the emission 
    arising of the X-ray heated face of the
    secondary Roche lobe. To match the orientation of Doppler
    maps usual in the literature we invert the sign convention of
    the $v_x$ and $v_y$ axes with respect to the $x$ and $y$ coordinates,
    respectively. The numerical calculation makes
    use of the fact that obviously the secondary star is totally
    opaque to X-rays and to optical line emission. The resulting
    self-occultation of the Roche lobe should provide a constraint on
    the  orbital inclination, but the endpoints  of the narrow emission line (NEL) eclipse
    are superimpose with emission from the stream. We therefore use the
    inclination determined for a given mass ratio from the eclipse width
    in the optical light curve.  The K-velocity and the width of the 
    secondary star emission feature can be used to
    estimate the relative Roche lobe size, and hence for the
    mass ratio. The maximum width of this feature is given as 2.45\AA
    \ by Schwope et al. (1997a), which unfortunately is not far from
    the spectral resolution (1.6\AA). 

    \begin{figure*}
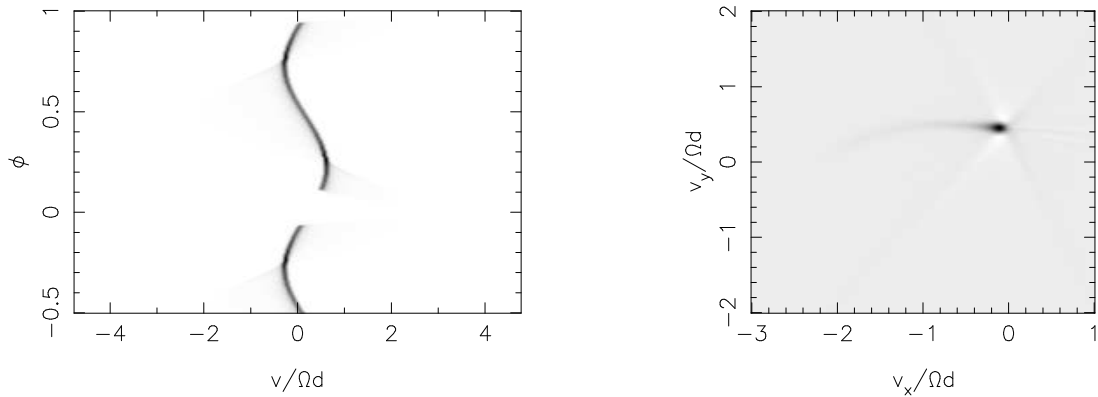
 \setlength{\unitlength}{1mm}
        \begin{picture}(170,70)
        \put(0,67){\includegraphics{images/fig5a.ps}}
        \put(90,67){\includegraphics{images/fig5b.ps}} \end{picture}
        \caption{Synthetic trailed spectrum of reprocessed line
        emission from the secondary star in assumption of a fully
        transparent stream in  X-rays and line. This is to be compared
        to the middle row of \fig~\ref{parts_fig}.  Doppler tomogram.}
        \label{thin_fig} \end{figure*}

\subsection{Emission of the horizontal stream}
    The horizontal stream emission is shown in the middle panel of
    \fig~\ref{spect}. Our model assumes that the horizontal stream is opaque
    to X-rays but transparent to the optical line photons.
    For phases smaller than
    approximately 0.55 we are viewing the irradiated side of the stream
    facing the X-ray source; in contrast, for
    phases larger that about 0.55 we are facing the un-irradiated side, however
    the observed emission in the \heii~4686\AA~line
    is just slightly lower by about 20\% 
    in the observations 
    (c.f. Schwope 1997a). So the horizontal stream
    cannot be opaque for both the X-ray irradiation and the optical
    line emission. Simulated tomograms (see \fig~\ref{thin_fig}) show
    that the stream cannot be fully transparent in both spectral
    regimes either. In that case the volume emissivity would be
    proportional to the square of the local density, which produces 
    too much emission close to the $L_1$-point.
    It has been observed that the Roche lobe is significantly shielded 
    from the X-ray irradiation by the accretion curtain, which is even 
    more dilute than the horizontal stream. So we suggest a horizontal 
    stream opaque to the ionising radiation but nearly fully
    transparent in the \heii~line so that we observe the irradiated 
    front side even when viewing through the stream.
    The low line opacity needed 
    seems plausible, given the large velocity gradients in
    the stream.

    These arguments make it plausible that the line radiation mainly
    stems from the skin layers of the stream, which probably is
    surrounded by some kind of hot X-ray heated corona. If we assume
    that the radiation reprocessing involves the same physical process
    as on the companion Roche lobe, then comparison of the relative
    fluxes allows an estimate of the stream width, which is
    proportional to $\epsilon$. Knowledge of
    $\epsilon$ in turn allows an estimate of the core temperature of
    the stream.

    Gaussian fitting of the high velocity component (HVC) emission shows clearly a fading  of
    the horizontal stream around phases 0 and 0.5 (gaps). This
    suggests a foreshortened optically thick emission. However the 
    above argument gives evidence for an optically thin stream transparent 
    to optical line photons. We therefore propose a thin X-ray heated
    region that is optically thick and reprocesses most of the ionising
    radiation whereas the remainder of the stream is optically thin.
    The resulting spectra are shown  in the middle panel of
    \fig~\ref{parts_fig}. We do not expect information from the gaps
    to be of great importance for the parameter fitting because of the
    rather poor signal to noise ratio at these phases. However,
    compared to the  observational data they occur at quite the right
    phase and velocity in our simulations. The rather narrow emission
    lines (knots) at $\varphi \approx 0.22$ and $\varphi \approx 0.72$
    occur when the orientation of the stream to the line of sight
    is such that the intrinsic velocity plus the orbital speed are
    virtually constant over the length of the stream. In addition at 
    $\varphi \approx 0$ occultation by the secondary star shields the
    stream from the observer.  Whereas the amplitude and the phases of
    the maxima of the horizontal  stream emission can be adjusted to
    the observational data by an appropriate  choice of the model
    parameters, the line profile of the high velocity wings provides a
    test to the correctness of our stripping model.

\subsection{Emission of the accretion curtain}
\begin{figure*}
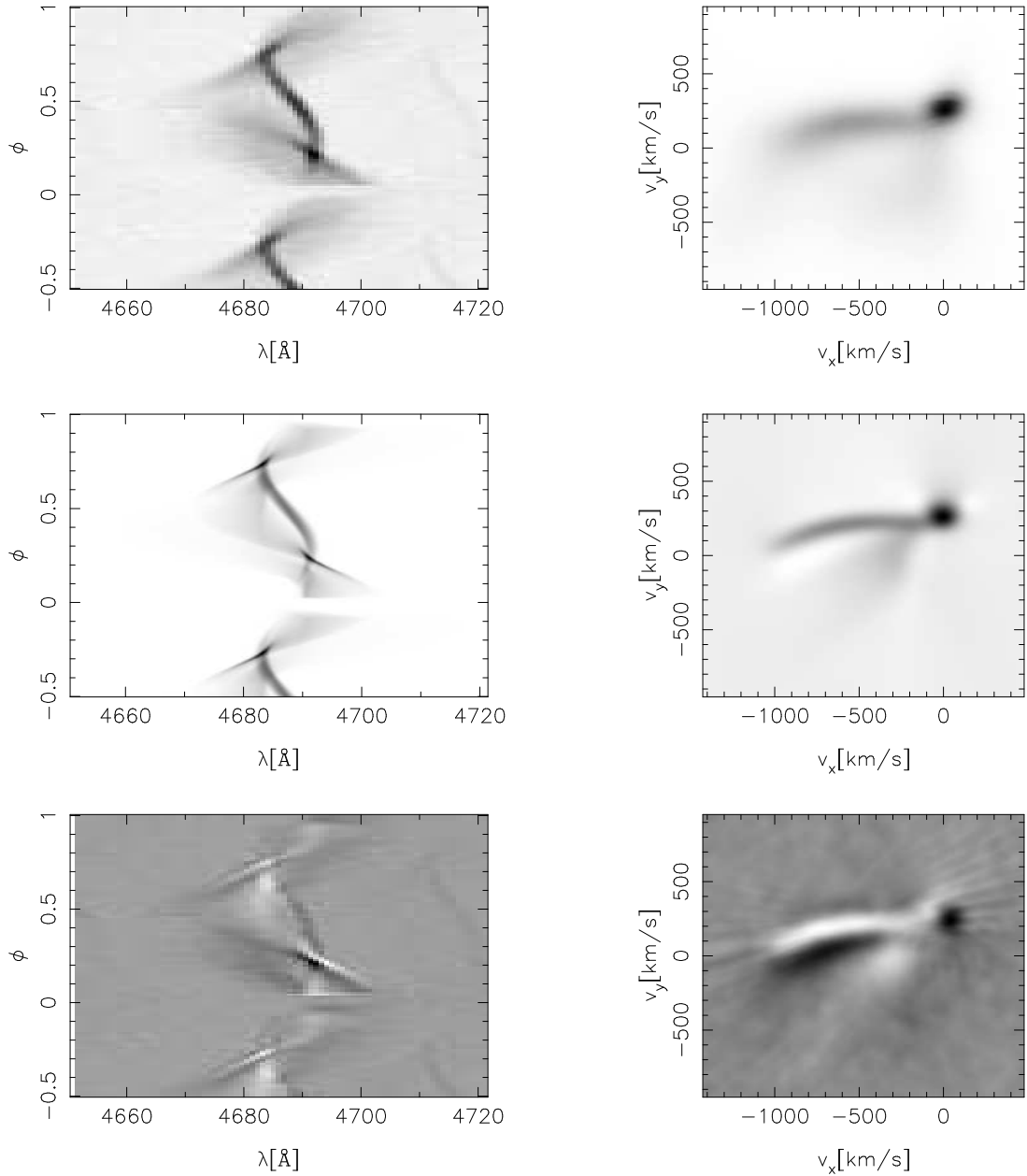

        \setlength{\unitlength}{1mm}
        \begin{picture}(170,170)
            \put(0,180){\includegraphics{images/fig6a.ps}}
            \put(90,180){\includegraphics{images/fig6b_MEM.ps}}
            \put(0,122){\includegraphics{images/fig6c.ps}}
            \put(90,122){\includegraphics{images/fig6d.ps}}
            \put(0,65){\includegraphics{images/fig6e.ps}}
            \put(90,65){\includegraphics{images/fig6f.ps}}
        \end{picture}
        \caption{Comparison between the observational data of HU~Aqr 
                in He II 4686 (upper row) and the model folded with 
                the spectral resolution (middle row), the residuals are
                shown in the bottom row. Displayed are the trailed 
                spectra (l.h.s.) and back-projected Doppler tomograms (r.h.s.), 
                apart from the top right panel where MEM tomography was used. 
               Model parameters from \tab~\ref{param_tab}.}
        \label{merge_fig}
    \end{figure*}

    The broad baseline component (BBC) emitted by the accretion curtain 
   (see bottom panel of \fig~\ref{parts_fig}) 
    appears very dim in the modelled trailed
    spectra and in the Doppler tomograms since it is distributed over
    a large range of velocities. 
    We assume the line emission is proportional to the intercepted X-rays.
    As for the horizontal
    stream the eclipse of the curtain is modelled with a spherical
    companion star  of the effective Roche radius. The brightest
    part of the curtain is the stagnation region in which the material
    enters the curtain when it couples to the magnetosphere.
    Emission from the curtain partially covers up the gaps of the
    horizontal stream emission in the trailed spectrum of the middle
    row of \fig~\ref{merge_fig}, in which the three components (Roche
    lobe, horizontal stream, and curtain) are merged.

\section{Testing the model}
\label{fit_sct}

    To adjust the 10 free parameters of the model, which are given in
    \tab~\ref{param_tab}, we have performed a $\chi^2$-fit to the
    HU~Aqr trailed spectra from August 1993, when the object was in
    its high accretion state. We focus on the
    4686\AA~\heii~line which has the best signal to noise ratio and
    shows the sharpest features of all lines in the observed spectral
    range from \mbox{4180 \AA} to 5070 \AA.  \fig~\ref{merge_fig} 
    compares the observational data and our fitted model,
    which was folded  by the observational resolution. The fitting 
    employed a simplex algorithm to adjust 8 parameters in order to
    minimise the  $\chi^2$ statistic for which we have used estimated
    standard deviations from a detector noise model. The orbital
    period $P$ is very well known from long-term light curves; also
    from the optical eclipse the orbital inclination $i$ of the system
    is given as a function of the mass ratio. So we have not varied
    these parameters in the fit. 

    There are certain features in the spectra which constrain
    particular parameters. With the period and mass ratio fixed the orbital separation $d$
    scales the wavelength axis. The mass fraction $\mu$ governs the
    width and amplitude of the Roche lobe emission. These two
    parameters hence are predominantly determined by the NEL feature, but
    the curvature of the horizontal stream also effects these
    parameters. The resulting Q~ratio of 3.9 lies in between the two
    values obtained by Schwope (1997a) from either of these features
    separately.  With $d$ and $\mu$ fixed, the length of the            
    horizontal stream is determined by the magnetic dipole parameters
    ${\cal M}$, $\varphi_d$ and $\vartheta_d$. However the magnetic field strength
    varies strongly with he distance from the white dwarf and only weakly with the 
    dipole orientation, thus the later cannot be well
    constrained alone from a fit to the trailed spectrogram. Therefore
    we employ the knowledge of the accretion spot position
    ($\varphi_s=40^o$, $\vartheta_s=26^o$), which is given by Schwope
    et al. (1997a), based on the  X-ray bright phase and cyclotron
    harmonics. We have used the mass radius relation for white dwarfs
    given by Pringle (1975) to constrain our variation of $\varphi_d$
    and $\vartheta_d$ to those values that are in agreement with the
    observed accretion spot.  The sound speed $\epsilon$ is a measure
    of the bulk temperature, which scales the geometrical width of
    the horizontal stream. Thus its total brightness provides the
    information to which $\epsilon$ is fitted.  The model also
    contains the relative brightness $I_c$ of the curtain and the
    rest wavelength $\lambda_0$ as free parameters.

    The error estimates given in \tab~\ref{param_tab} have been
    obtained by bootstrapping the original observation.  On each pixel
    a random signal was added with the detector noise as standard
    deviation. We have repeated the $\chi^2$-fit starting with the
    same initial values for 20 different samples. The errors in
    \tab~\ref{param_tab}  are $1 \sigma$ intervals of the fitted
    parameters and consequently are only a measure of the uncertainty
    due to the  noise in the data. They certainly do not account for
    systematic discrepancies of the model from reality   (see
    \sct~\ref{discus}).

    \begin{table}
        \caption{Model parameters obtained from $\chi^2$-fitting; these are used for the figures 
                 shown in this article. The quantities where no error is given have not been varied 
                 in the $\chi^2$-fit.}
        \begin{center}
        \begin{tabular}{lccc}
           \hline
           parameter              & symbol          &  value        & error \\
           \hline
           orbital period         & $P$             & $0.\!^d086820446$          & --\\ 
           orbital inclination    & $i$             & $84.\!^o0$                 & --     \\
           orbital separation     & $d$             & $5.663 \! \cdot \! 10^{10} {\rm cm}$ & $5.6 \! \cdot \! 10^{8} {\rm cm}$\\
           mass fraction          & $\mu$           & $0.7950$                   & $0.0027$\\    
           magn.~momentum      & $\epsilon {\cal M}$ & $0.0771$                   & $0.0035$\\ 
           dipole azimuth         & $\varphi_d$     & $37.\!^o8$                 & $1.\!^o6$\\
           dipole co-latitude     & $\vartheta_d$   & $12.\!^o4$                 & $0.\!^o3$\\
           sound speed            & $\epsilon$      & $0.0355$                & $0.0016$\\ 
           rest wavelength        & $\lambda_0$     &  4685.7\AA                & 0.1\AA\\
           rel.~curtain intens. & $I_c$    &   $0.0154$                   & $0.0023$\\
           \hline
        \end{tabular}
        \end{center}
        \label{param_tab}
   \end{table}

    From the fitted model parameters we are able to derive a couple of
    physical parameters of the HU~Aqr system. With the observed
    magnetic field at the accretion spot of 35 MG we are able to scale
    our intrinsic units from \eqn~(\ref{units}) to physical
    units. This enables us to determine the total mass accretion rate
    as $\dot{M} = 2.9 \times 10^{-10}\, {\rm M}_\odot/{\rm yr} \,(B/35 {\rm MG})^2 (\epsilon/0.0355)^2$. 
    This value depends
    strongly on our assumption of identical effectivity of
    radiation reprocessing on the companion star surface and the
    horizontal stream. Here also the choice of the stripping condition
    and the assumption of a quasi-steady flow are very
    essential. Nevertheless the derived $\dot{M}$ is not extremely far
    from $\dot{M} \approx 0.5 \times 10^{-10}\, {\rm M}_\odot/{\rm yr}$ estimated
    from the total luminosity of an average CV just below the period
    gap (Patterson 1984). The accretion rate in the model is lowered
    by a factor of 4 when either the radiation reprocessing in the
    horizontal stream is twice as effective as on the Roche lobe, or
    if matter coupling to the magnetosphere needs a magnetic
    field twice as large as assumed. 

    In the optical light curve a pre-eclipse dip centred at
    phase $\phi=-0.118$ has been observed; it has a symmetric 
    distribution that can well be
    approximated by a Gaussian profile with a dispersion of $0.015$.
    Our model cannot account for this dispersion, which is presumably
    a consequence of the matter being  broken up into blobs of
    different sizes, hence the threading points vary over a range of 
    locations. The good agreement of the modelled turn-off point
    at phase  $\phi=-0.117$ shows that the horizontal stream is free
    falling in a good approximation. So the geometrical location of
    the coupling region and the fitted location in velocity space
    coincide.

    The model also gives strong evidence that the companion star is on
    the main sequence. The radius of a main sequence star with a mass
    of  $0.19 {\rm M}_\odot$ is given by Neece (1984) as $0.23 {\rm R}_\odot$
    which agrees very well with the effective radius of the Roche lobe
    provided by Eggleton (1983) which comes out to be $0.22 {\rm R}_\odot$
    for our values of $\mu$ and $d$. We do not have to assume a mass
    radius relation for this result.

    At the coupling point we have a mass density of $7.6 \times 10^{-11}
    {\rm g/cm}^3$ or, for a pure hydrogen gas, a number density of
    $n=4.2 \times 10^{16}{\rm cm}^{-3}$, which results in a Debye length of
    $\lambda_d=4.1 \times 10^{-3}{\rm cm}$. Hence we have an  ideal plasma with
    a plasma parameter $\Gamma := 1/(n \lambda_d^3) = 3 \times 10^{-10}$. 

   \begin{table}
        \begin{center}
        \caption{Some physical quantities of HU~Aqr as derived from the model parameters 
        given in \tab~\ref{param_tab}.}
        \begin{tabular}{lcc}
           \hline
           quantity              & symbol                 &  value       \\
           \hline
           mass ratio             & $Q$                    & $3.9$              \\    
           dip angle              & $\phi_{\mbox{\tiny dip}}$              &  $  -0.117  $    \\
           primary mass           & $M_{\mbox{\tiny wd}}$  & $ 0.75 {\rm M}_\odot$                \\    
           secondary mass         & $M_{\mbox{\tiny rd}}$  & $ 0.19 {\rm M}_\odot$               \\    
           white dwarf radius     & $R_{\mbox{\tiny wd}}$  & $ 7.3\!\cdot\!10^8 {\rm cm}$                \\    
           secondary Roche radius & $R_{\mbox{\tiny rd}}$  & $ 1.5\!\cdot\!10^{10} {\rm cm}$                 \\    
           bulk temperature     & $T$              &  $ 17\!\cdot\!10^3 {\rm K}  $ \\
           total mass flow        & $\dot{M}$              &  $  2.9\!\cdot\!10^{-10}\, {\rm M}_\odot/{\rm yr}   $    \\ 
           coupling density     & $n$              &  $ 7.6\cdot10^{-11} {\rm g/cm}^3  $ \\
           \hline
        \end{tabular}
        \end{center}
        \label{deriv_tab}
   \end{table}

\section{Discussion}
    \label{discus}

    In the comparison between our model and  the observational data of
    \fig~\ref{merge_fig}  five details differ significantly: (1) At
    the location of the Roche lobe  feature the residual tomogram shows a
    dark spot at positive $v_x$ and a bright region towards  negative
    $v_x$. This is a consequence of the missing shielding of X-ray
    flux by the curtain in our model.    (2) The horizontal stream in
    the observation is clearly shifted in direction of smaller
    $v_y$. This is not  a tomographic artefact as it is also clear from
    the residuals in the trailed spectrum near the  knot at
    $\phi=0.22$ that the simulation and observation are shifted
    against each other. The reason  for this might be a magnetic drag
    force that acts transversely on the horizontal stream before the
    matter is totally coupled to the field.   (3) The
    curtain emission seems to be too faint at low velocities in the
    model spectra and also ranges out  to too high velocities. Also
    the curtain is located at lower $v_y$ in the model than in the 
    tomogram  of the observed data.  This could be explained either by
    a reduction of radiation reprocessing when the curtain gets closer
    to the  white dwarf, or by the importance of pressure effects when
    the curtain is funneled towards the  accretion spot, or as well by
    a distortion of the dipole geometry. There also is a
    no\-ti\-ca\-ble difference   in the shape of the eclipse, so it
    might be suggested that the reason lies in the geometry of the
    curtain flow.  (4) The  observed tomogram  shows faint but
    significant emission in the region roughly between
    $(v_x,v_y)=(-900,0)$ to $(-1200,-600)$, which cannot be explained
    in our model. A plausible speculation on the origin of this
    emission could be larger blobs in the horizontal stream that are
    sufficiently massive to survive the stripping process but feel a
    strong drag force by the magnetic field that forces them to leave
    the ballistic trajectory.  (5) Even though folded with the
    observational resolution the widths of all modelled stream
    features are too   small in both the trailed spectra and the
    Doppler tomograms. This could be caused by an underestimate of the
    observational  resolution, or an intrinsic line
    broadening. Thermal broadening of the bulk temperature is already 
    included in the model, the gas is much too dilute  for pressure 
    broadening, and the field strength is
    much to low for a noticeable Zeeman effect. However the layer in
    which the radiation reprocessing occurs may be heated far above
    the bulk temperature. This hot corona-like gas around the stream 
    could be the source of the large observed line width. Also broadening 
    by turbulent gas motion might be important.  

    We have developed a predictive model for the stripping of the
    accretion  stream and the line emission in AM Her stars in general and
    applied it to  spectral observations of the eclipsing polar HU~Aqr in
    its high  accretion state. The main observed features, including the
    length of the horizontal stream and the main emission line components,
    are well reproduced  by our model. A $\chi^2$-fit of synthesised to
    observed trailed spectrograms has uncovered a number of physical
    parameters of the binary system (mass flow rate, stream temperature,
    stream density, orientation of the magnetic field, mass ratio
    etc.). The  detailed comparison between observation and simulation
    (c.f.~\fig~\ref{merge_fig}) shows some systematic deviations which may
    be overcome by further improvements of the model. 
    A possible cure for the above problems could be a numerical
    calculation of the field distortion in presence of a highly
    conducting plasma and a more detailed investigation on the flow
    instabilities. However on account of the enormous Reynolds numbers
    of the flow a fully self-consistent numerical model seems not
    feasible to us. A further improvement for the model would be a
    more precise calculation of the radiation reprocessing rather that
    just assuming limiting cases for the opacity, especially the
    velocity gradients in the stream could significantly alter the
    characteristics of the line radiation.  However the overall agreement
    between observational data and the proposed model based on rather
    simplified physics seems remarkably good to us.

\section*{Acknowledgements}
    The authors would like to thank Danny Steeghs for providing a 
    MEM tomogram of HU~Aqr, and Rick Hessman, Thomas Neukirch,
    Moira Jardine and Miguel Ferreira for very informative
    discussions.  We gratefully acknowledge the Cormack Bequest
    Scholarship of the Royal Society of Edinburgh. Without its help 
    and the hospitality of the AIP in Potsdam this project could 
    not have been completed. This work has also been
    supported in part by the Deutsches Zentrum f\"ur Luft- und
    Raumfahrt (DLR, former DARA) under grant 50 OR 9403 5.

\bibliographystyle{mn}
\bibliography{}

\appendix

\section{Power series for the horizontal stream at $L_1$ }
    \label{series_app}
\subsection*{Vertical scale}
    In their classical paper on gas dynamics in semi-detached binaries 
    Lubow \& Shu (1975) assume that thermal equilibrium prevails in 
    $z$-direction. In a later publication (1976) they give also a closed 
    set of differential equations for the vertical structure of the stream:
    \begin{eqnarray}
        u_{s0} \, \frac{d \chi}{ds} + 2\nu \chi &=& 0  \\
        u_{s0} \, \frac{d \nu}{ds} + \nu^2 - \chi 
             + \Big( \frac{\partial^2 \Phi}{\partial z^2} \Big)_0 &=& 0 \\
        \chi(0) = A \;,  \quad   \nu(0) &=& 0
    \end{eqnarray}
    Therein is $1/\sqrt{\chi}\,$  the dispersion of the vertical Gaussian, 
    $\nu=\frac{\partial u_z}{\partial z}$ the divergence of the 
    $z$-velocity and $\Big( \frac{\partial^2 \Phi}{\partial z^2} \Big)_0$ 
    the second derivative of the Roche potential taken at the stream centre.

    $L_1$ is a singular point of these equations since energy conservation 
    provides $u_{s0} = \lambda_1 s +  O(s^2)$ and hence $u_{s0}$ is 
    vanishing at initial point $s=0$. The similarity to Euler's equation 
    suggests the `Ansatz' as a fractional power series
    \begin{eqnarray}
        \chi = A + \sum_{k=0}^{\infty} C_k \, s^{k+c}\,,  \quad  
         \nu = \sum_{k=0}^{\infty} N_k \, s^{k+b}
    \end{eqnarray} 
    for some $b>0$, $c>0$ and sets $\{C_k\}_k$ and$\{N_k\}_k$ with 
    $C_0 \not = 0$ and  $N_0 \not = 0$, which provides uniqueness of the 
    solution. After substitution and use of  Cauchy's theorem  the 
    $s^b$-term in the $\chi$-equation and the $s^1$-term in the 
    $\chi$-equation provide $b=c=1$, using the non-vanishing of $C_0$ 
    and $N_0$. This proves that the solution we are looking for is 
    even analytical at $L_1$, which has not been obvious a priory.

    Knowing $c$ and $b$ we can compare the terms in $s^k$ and obtain 
    to first order
    \begin{eqnarray}
        \chi = A +  \frac{2 A \Phi_{zzs}}{\lambda_1^2 + 2A}\,s + O(s^2) \\
        \nu = \frac{\lambda_1 \Phi_{zzs}}{\lambda_1^2 + 2A}\,s + O(s^2)
    \end{eqnarray} 
    where, with the white dwarf mass fraction $\mu$, we have been using 
    the definitions
    \begin{eqnarray}
        A &:=& \frac{\mu}{|x_{\mbox{\tiny L}_1}-1+\mu|^3} \,
              +\, \frac{1-\mu}{|x_{\mbox{\tiny L}_1}+\mu|^3}   \\
        \lambda_1^2 &:=& \frac{1}{2} \Big( A-2+\sqrt{9A^2-8A} \, \Big)  \\
        \Phi_{zzs} &:=& \Big(\frac{\partial^3 \Phi}{\partial^2 z \partial s}
                   \Big)_{{\mbox{\tiny L}_1}} 
    \end{eqnarray}

\subsection*{Horizontal scale}
    Let  $ 1/\sqrt{\gamma}\,$ be the horizontal dispersion of the
    Gaussian density profile of the stream, $\beta=\frac{\partial
    u_n}{\partial n}$ the divergence of the $n$-velocity and $K=1/R$
    the curvature of the ballistic trajectory. Then the differential
    equations for the horizontal stream width are, as derived by Lubow
    \& Shu (1975):
    \begin{equation}
        u_{s0} \, \frac{d \gamma}{ds} + 2\beta \gamma = 0 
    \end{equation}
    \begin{equation}
        u_{s0} \, \frac{d \beta}{ds} + \beta^2 -\gamma = 
                 (2 - 3Ku_{s0})Ku_{s0} - \Big(\frac{\partial^2 \Phi}
                {\partial n^2}\Big)_0 - 4  \\
    \end{equation}
    \begin{equation}
        \beta(0) = 0  \,  ,  \quad  \gamma(0) = \lambda_{1}^2 -A+3 =: 
               \lambda_{3}^2 \;,
    \end{equation}
    where we have already substituted the analytical integrals found
    by Lubow \& Shu. The singularity of these equations is of the same
    kind as for the vertical structure, hence we are claiming again
    that $\gamma$ and $\beta$ can be written as fractional power
    series. A somewhat similar but more extensive algebra shows then 
    that the solution of the horizontal equations is also analytical. 
    As first order solution we obtain
    \begin{eqnarray}
        \gamma &=& \lambda_{3}^2 + \frac{(12K_0\lambda_{1}-\Phi_{nns})
                \lambda_{3}^2}{2\lambda_{3}^2+\lambda_{1}^2}\,s + O(s^2) \\
        \beta &=& \frac{(6\lambda_{1}-\Phi_{nns})\lambda_{1}}
               {2\lambda_{3}^2+\lambda_{1}^2}\,s + O(s^2) \,,
    \end{eqnarray} 
    where $K_0$ is the curvature at the $L_1$-point and
    \begin{eqnarray}
        \lambda_3^2&:=&\lambda_1^2-A+2\\
        \Phi_{nns} &:=& \Big(\frac{\partial^3 \Phi}{\partial^2 n 
               \partial s}\Big)_{{\mbox{\tiny L}_1}} 
    \end{eqnarray}

\section{Magnetic focusing in the curtain plane}
    \label{focusing}
    We  now consider the magnetic focusing  in the curtain plane. Be $P$ a
    fixed point with co-latitude $\theta$ on an arbitrary field line
    $(r_m, \psi)$. On a neighbouring field line $(r_m+d r_m, \psi + d
    \psi)$ a point $Q$ with the co-latitude $\theta + d \theta$ has a
    distance $dl$ from $P$ which is to first order in the locally
    Euclidean metric
    \begin{equation}
        (d l)^2 = (d r)^2 + (r d\theta)^2 + (r \sin\theta d\psi)^2
        \label{metric}
    \end{equation}   
    where
    \begin{eqnarray}
        r &=& r_m \sin^2\theta \\
        d r &=& (r_m+d r_m) \sin^2(\theta + d\theta) - r = \nonumber \\
            &=& (2 r_m \sin\theta \cos\theta ) d\theta + (\sin^2\theta) d r_m 
    \end{eqnarray}   
    The condition for a critical point is given by \mbox{$\frac{\partial}
    {\partial (d\theta)} (d l)^2 = 0$} and yields as a unique solution
    \begin{equation}
        d\theta = -2 \, \frac{\cos\theta \sin\theta}
                 {r_m(1+3\cos^2\theta)}\; dr_m \, ,
    \end{equation}   
    which, if $\theta>-d\theta$, actually provides by substitution into 
    \eqn~(\ref{metric}) the real minimum distance $d l$ between P and 
    any point $Q(\theta+d\theta)$ on a neighbouring field line:
    \begin{eqnarray}
       (d l)^2 &=& h^2_{r_m} (d r_m)^2 + h^2_{\psi}(d\psi)^2  \\
      h^2_{r_m} &:=& \frac{ 1-\cos^6\theta+3\cos^4\theta-3cos^2\theta}
          {1+3\cos^2\theta} \label{h_r_m}\\
      h^2_{\psi} &:=& \frac{r_m \sin^2\theta(1+\cos^2\theta) }
          {1+3\cos^2\theta} \label{h_psi}  
    \end{eqnarray}   

    \label{lastpage}
\end{document}